\documentstyle[11pt,epsf,psfig]{article}

\newskip\humongous \humongous=0pt plus 1000pt minus 100pt
\def\caja{\mathsurround=0pt}
\def\eqalign#1{\,\vcenter{\openup1\jot \caja
       \ialign{\strut \hfil$\displaystyle{##}$&$
        \displaystyle{{}##}$\hfil\crcr#1\crcr}}\,}
\newif\ifdtup

\newcounter{eqnumber}
\renewcommand{\theeqnumber}{\arabic{eqnumber}}
\def\equn{
\refstepcounter{eqnumber}
\eqno({\rm \theeqnumber})
}

\def\isjunk#1{}

\def\eqn#1{eq.~(\ref{#1})}
\def\Eqn#1{Equation~(\ref{#1})}
\def\eqns#1#2{eqs.~(\ref{#1}) and~(\ref{#2})}

\def\fig#1{fig.~{\ref{#1}}}

\def\tr{\mathop{\rm tr}\nolimits}

\def\qcd{{\rm gauge\!}}

\newbox\charbox
\newbox\slabox
\def\s#1{{      
        \setbox\charbox=\hbox{$#1$}
        \setbox\slabox=\hbox{$/$}
        \dimen\charbox=\ht\slabox
        \advance\dimen\charbox by -\dp\slabox
        \advance\dimen\charbox by -\ht\charbox
        \advance\dimen\charbox by \dp\charbox
        \divide\dimen\charbox by 2
        \raise-\dimen\charbox\hbox to \wd\charbox{\hss/\hss}
        \llap{$#1$}
}}

\def\spa#1.#2{\left\langle#1\,#2\right\rangle}
\def\spb#1.#2{\left[#1\,#2\right]}
\def\spab#1.#2.#3{\langle\mskip-1mu{#1}^- 
                  | #2 | {#3}^-\mskip-1mu\rangle}
\def\spba#1.#2.#3{\langle\mskip-1mu{#1}^+ 
                  | #2 | {#3}^+\mskip-1mu\rangle}
\def\spaa#1.#2.#3{\langle\mskip-1mu{#1}^-
                  | #2 | {#3}^+\mskip-1mu\rangle}
\def\spbb#1.#2.#3{[\mskip-1mu{#1}^+
                  | #2 | {#3}^-\mskip-1mu]}
\def\lor#1.#2{\left(#1\,#2\right)}

\catcode`@=11  

\def\eps{\epsilon}

\def\A{{\cal A}}
\def\M{{\cal M}}

\def\Perm{{\cal P}}

\def\pol{\varepsilon}

\def\la{\langle}
\def\ra{\rangle}
\def\tree{{\rm tree}}
\def\oneloop{{1 \mbox{-} \rm loop}}

\def\ksl{\s{k}}
\def\Ksl{\s{K}}

\def\Split{\mathop{\rm Split}\nolimits }

\def\SplitGrav{\mathop{\rm Split}\nolimits^{\rm gravity} }
\def\SoftGrav{\mathop{\cal S} \nolimits}

\def\del{\partial}

\begin{document}

\begin{titlepage}

\begin{flushright}

hep-th/9809160 \hfill SLAC--PUB--7953\\
UCLA/98/TEP/36\\
UFIFT-HEP-98-24\\
September, 1998\\
\end{flushright}

\vskip 2.cm

\begin{center}
\begin{Large}
{\bf One-Loop $n$-Point Helicity Amplitudes in (Self-Dual) Gravity}
\end{Large}

\vskip 2.cm

{\large Z. Bern$^{\star,1}$, L. Dixon$^{\dagger,2}$, 
M. Perelstein$^{\dagger,2}$ and J.S. Rozowsky$^{\ddagger,3}$}

\vskip 0.5cm

$^\star${\it Department of Physics,
University of California at Los Angeles,
Los Angeles,  CA 90095-1547}

\vskip .3cm

$^\dagger${\it Stanford Linear Accelerator Center,
Stanford University, Stanford, CA 94309}

\vskip .3cm
$^\ddagger${\it Institute for Fundamental Theory, Department of Physics, 
\vskip .01cm
University of Florida, Gainesville, FL 32611}

\vskip .3cm
\end{center}

\begin{abstract}
We present an ansatz for all one-loop amplitudes in pure Einstein gravity
for which the $n$ external gravitons have the same outgoing helicity.
These loop amplitudes, which are rational functions of the momenta, also
arise in the quantization of self-dual gravity in four-dimensional
Minkowski space.  Our ansatz agrees with explicit computations via
$D$-dimensional unitarity cuts for $n \leq 6$.  It also has the expected
analytic behavior, for all $n$, as a graviton becomes soft, and as two
momenta become collinear.  The gravity results are closely related to
analogous amplitudes in (self-dual) Yang-Mills theory.
\end{abstract}

\vskip 1cm
\begin{center}
{\sl Submitted to Physics Letters B}
\end{center}

\vfill
\noindent\hrule width 3.6in\hfil\break
${}^{1}$Research supported by the US Department of Energy
under grant DE-FG03-91ER40662.\hfil\break
${}^{2}$Research supported by the US Department of Energy under grant 
DE-AC03-76SF00515.\hfil\break
${}^{3}$Research supported by the US Department of Energy under grants 
DE-FG03-91ER40662 and DE-FG02-97ER41029.\hfil\break
\end{titlepage}

\baselineskip 16pt


\section{Introduction}
\label{IntroSection}

Gravity and Yang-Mills theory have many similarities at the classical
level.  Both theories are nonlinear, and possess the respective local
symmetries of general coordinate invariance and non-abelian gauge
invariance.  In the weak field limit, they both admit plane-wave solutions
corresponding in the quantum theory to massless particles, gravitons and
gluons.  Also, a rich set of exact solutions is known for the field
equations of each theory restricted to self-dual configurations; for
example, the multi-instanton solutions of Yang-Mills theory on
$S^4$~\cite{Atiyah}, and various gravitational
instantons~\cite{GravInstantons}.

At the quantum level, however, the two theories behave quite differently.
The dimensionless non-abelian coupling of pure gauge theory becomes
logarithmically strong in the infrared, leading to confinement of colored
quanta, whereas in the ultraviolet the theory is renormalizable and in
fact, asymptotically free.  On the other hand, the dimensionful nature of
Newton's constant means that the infrared behavior of gravity is
well-described by the classical limit, but also implies that its
ultraviolet behavior is nonrenormalizable by power-counting arguments.
Thus, at the quantum level gravity should presumably be regarded as only
an effective low-energy limit of some more fundamental theory, such as
string theory.

It is nevertheless interesting to examine more carefully the quantum
(loop) behavior of gravity, and its connection with gauge theory.  The
scattering amplitudes of gravitons and of gluons that have been
investigated to date have proven to be quite closely related.  Classical
(tree-level) amplitudes for gravity obey a `squaring relation', derived by
Kawai, Lewellen and Tye (KLT) from string theory, in which each graviton
amplitude is given, roughly speaking, by the sums of products of pairs of
gluon amplitudes~\cite{KLT,BGK}.  These $n$-point tree-level KLT
relations, in conjunction with the unitarity of the $S$-matrix, have
recently led to similar relations between four-point amplitudes at the
multi-loop level, for the maximally supersymmetric versions of gravity and
gauge theory, $N=8$ supergravity and $N=4$ super-Yang-Mills
theory~\cite{SusyEight}.  Such relations have led to an improved
understanding of the ultraviolet behavior of $N=8$ supergravity in various
dimensions.

In this letter we present an ansatz for the one-loop amplitudes in pure
Einstein gravity (${\cal L} = -{2\over\kappa^2} \sqrt{-g} R$,
$\kappa=\sqrt{32\pi G_N}$) with an arbitrary number of external gravitons,
all having positive helicity.\footnote{ Our crossing-symmetric convention
is to label outgoing states by their helicity, and incoming states by the
reversed helicity (i.e. the helicity they would have if crossed into the
final state).}  These `all-plus' gravity amplitudes are closely related to
the previously computed one-loop all-plus gauge
amplitudes~\cite{AllPlus,MahlonAllPlus}.  Both sets of amplitudes
correspond to self-dual configurations of the field strengths, obeying
respectively $R_{\mu\nu\rho\sigma} = {i\over
2}\eps_{\mu\nu}{}^{\alpha\beta} R_{\alpha\beta \rho\sigma}$ and
$F_{\mu\nu} = {i\over 2} \eps_{\mu\nu}{}^{\alpha\beta} F_{\alpha\beta}$,
with $\eps_{0123} = +1$.  Self-dual gravity (SDG)~\cite{SDG} and self-dual
Yang-Mills theory (SDYM)~\cite{SDYMActions,SDYMReview} have attracted
attention through their connection with integrable models, twistor theory
and $N=2$ string
theory~\cite{SDYMReview,Integrable,NTwoStringOld,NTwoStringOV,
NTwoOpenHetString,BerkVafa}.

Strictly speaking, these self-dual theories are only defined as full
quantum theories in four-dimensional space-times with an even number of
time dimensions, i.e. signatures $(0,4)$ (Euclidean space) and $(2,2)$
(complex time), where the above self-dual constraint lacks an `$i$' and is
compatible with the reality of the fields.  However, a self-dual sector of
the full theory can be defined in signature $(1,3)$ (four-dimensional
Minkowski space).  At the linearized level, the classical solutions are
circularly polarized plane waves, corresponding to superpositions of
states of identical helicity (which may have different momenta).  For such
solutions in gauge theory, the complexified chromo-electric and
chromo-magnetic fields satisfy $E_j^a = - i B_j^a$, where $j$ is a spatial
index and $a$ an adjoint gauge index.  The gravitational analogs of
$E_j^a$ and $B_j^a$ are $E_{jk} \equiv R_{j0k0}$ and $B_{jk} \equiv
{1\over2} \epsilon_{j0}{}^{mn} R_{mnk0}$; they similarly satisfy 
$E_{jk} = - i B_{jk}$ for circularly polarized gravitational waves.

It is nontrivial that the connection between positive helicity amplitudes
and self-duality survives the full nonlinear interactions, as demonstrated
for gauge theory by Duff and Isham~\cite{DuffIsham}.  More recently,
several authors~\cite{Bardeen,Selivanov,Cangemi} have shown that various
SDYM Lagrangians~\cite{SDYMActions} (whose classical field equations all
solve the SDYM constraints) could be used to compute the tree-level matrix
elements for an off-shell gauge current to produce $(n-1)$
identical-helicity on-shell gluons in the full gauge theory.  The
tree-level scattering amplitudes that result from putting the current
on-shell actually vanish for all $n$ by a supersymmetry Ward identity
(SWI)~\cite{SWI},
$$
\M_n^{\rm SUSY}(1^\pm,2^+,\ldots,n^+) = 0,
\equn\label{vanishSWI}
$$   
where the external states, labeled by their helicity, may be
either gauge bosons or gravitons.  Eqn.~(\ref{vanishSWI}) applies 
also to tree amplitudes for nonsupersymmetric Yang-Mills theory, and
gravity, because fermionic superpartners never contribute to tree graphs 
for bosonic amplitudes.

At one loop, identical-helicity amplitudes in non-supersymmetric theories
do not vanish.  They are known explicitly for pure Yang-Mills theory for all 
$n$~\cite{AllPlus,MahlonAllPlus}.
Cangemi~\cite{Cangemi} and Chalmers and Siegel (CS)~\cite{ChalmersSiegel}
showed that these loop amplitudes can also be computed from various 
SDYM Lagrangians (up to an overall factor of two).  The result for the 
CS Lagrangian,
$$
{\cal L}_{\rm CS}^{\rm SDYM}
 = \tr \Bigl[ \bar\phi \Bigl( \del^2 \phi 
   + ig(\del^{\alpha}{}_{\dot{+}}\phi)
       (\del_{\alpha\dot{+}}\phi) \Bigr) \Bigr] \,,
\equn\label{CSSDYM}
$$
where two-component spinor notation has been used, agrees with the
pure-Yang-Mills result including the factor of two.  The
action~(\ref{CSSDYM}) is obtained by truncating the light-cone action 
for $N=4$ supersymmetric Yang-Mills theory~\cite{LCNFour}; 
the `$\dot{+}$' spinor
is defined with respect to the light-cone direction.  The 
corresponding truncation of the $N=8$ supergravity action contained in
ref.~\cite{SiegelSelfDualSugra} gives an action for self-dual 
gravity~\cite{ChalmersSiegel,CSUnpublished},
$$
{\cal L}_{\rm CS}^{\rm SDG}
 =  \bar\phi \Bigl( \del^2 \phi 
   + {\kappa\over2}(\del^{\alpha}{}_{\dot{+}}\del^{\beta}{}_{\dot{+}}\phi)
                   (\del_{\alpha\dot{+}}\del_{\beta\dot{+}}\phi) \Bigr) \,.
\equn\label{CSSDG}
$$
Note that the kinematic structure of the 3-point vertex in \eqn{CSSDG}
is just the square of that in \eqn{CSSDYM}, a feature which is consistent 
with the KLT relations.

For both \eqns{CSSDYM}{CSSDG}, $\bar\phi$ is the loop counting
parameter:  Tree amplitudes (which all vanish on-shell) have one external
$\bar\phi$, one-loop amplitudes have no external $\bar\phi$'s, and there
are no $(l>1)$-loop amplitudes.\footnote{This does not mean that
identical-helicity amplitudes vanish in full gauge theory, or gravity, for
$l>1$, just that the connection with self-dual theories breaks down
at that point.}  Thus the one-loop all-plus amplitudes for Yang-Mills theory
(gravity) are the {\it only} nonvanishing scattering amplitudes in the 
CS version of SDYM (SDG).

To construct an ansatz for the $n$-point one-loop all-plus graviton
amplitudes, $\M_n^\oneloop(1^+,2^+,\ldots,n^+)$, we have used a
combination of explicit computation (for $n=4,5,6$), and general analytic
properties (to infer the all-$n$ result).  The analytic properties are
very similar to those of the all-plus gauge
amplitudes~\cite{AllPlus,MahlonAllPlus}.  First of all, the unitarity cuts
for the amplitudes are identically zero in four dimensions: Each one-loop
cut is a product of two tree amplitudes, one on each side of the cut; but
every possible assignment of helicity to the two gravitons crossing the
cut leads to the vanishing of at least one of the two tree amplitudes, via
\eqn{vanishSWI}.  Similar reasoning shows that the one-loop all-plus
amplitudes do not contain multi-particle poles, of the form
$1/(k_{i_1}+k_{i_2}+\cdots+k_{i_m})^2$ with $m>2$.  The only permitted
kinematic singularities are those where one external momentum becomes
soft, or two momenta become collinear.  These singularities have a known
universal form, which will be described in more detail below.  Finally,
the loop-momentum integration does not generate any infrared nor
ultraviolet divergences.  In summary, $\M_n^\oneloop(1^+,2^+,\ldots,n^+)$
is a finite rational function of the momenta, totally symmetric in the $n$
arguments, with only soft and collinear singularities.

For the explicit computation, we first used the SWI~(\ref{vanishSWI}) to
replace the one-loop amplitude with a graviton in the loop by that with a
massless scalar in the loop~\cite{GZ}.  We then calculated the cuts for
this scalar loop in an arbitary dimension $D$, where they are
nonvanishing.  From the $D$-dimensional cut information, one can extract
the $D=4$ amplitude~\cite{Review}.  Further details of this
computation, and an intriguing relation between the $D$-dimensional
all-plus amplitudes and certain $N=8$ supergravity amplitudes, may be
found in ref.~\cite{AllnSusy}.

From the form of the all-plus amplitudes for $n=4,5,6$, and particularly
their factorization properties as a graviton momentum becomes soft, we
have arrived at an ansatz for the remaining amplitudes, $n\ge 7$.  As an
additional check, we have verified that the amplitudes factorize properly
as two gravitons become collinear.


\section{The Ansatz and Its Soft Limits}
\label{AnsatzSection}

To motivate the ansatz for the one-loop all-plus graviton
amplitudes, we briefly describe the only other known nontrivial
infinite sequence of graviton amplitudes.  These are the tree amplitudes
with maximal helicity violation (MHV), consistent with \eqn{vanishSWI},
for which exactly two gravitons have opposite helicity from the 
remaining $n-2$.  Berends, Giele and Kuijf (BGK)\cite{BGK} found the 
following compact expression,\footnote{Our overall phase conventions 
differ from those of ref.~\cite{BGK} by a `$-i$'.}
$$
\eqalign{
\M_n^\tree(1^-, 2^-, 3^+, \ldots, n^+) &= 
-i \left({\kappa\over2}\right)^{n-2} \, {\spa1.2}^8 \cr
& \hskip-3cm \times \Biggl[ 
 { \spb{1}.{2} \spb{n-2}.{\ n-1} \over \spa{1}.{\ n-1} \, N(n) }
 \biggl( \, \prod_{i=1}^{n-3} \prod_{j=i+2}^{n-1} \spa{i}.{j} \biggr)
   \prod_{l=3}^{n-3} \Bigl( - \spab{n}.{\Ksl_{l+1,n-1}}.{l} \Bigr)
\ +\ \Perm(2,3,\ldots,n-2) \Biggr] \,, \cr}
\equn\label{BGKMHV}
$$
where
$$
N(n) \equiv \prod_{i=1}^{n-1} \prod_{j=i+1}^n \spa{i}.{j} \,,
\equn\label{NnDef}
$$
$K^\mu_{i,j} \equiv \sum_{s=i}^{j} k_s^\mu$, and $+\Perm(M)$ instructs one
to sum the quantity inside the brackets over all permutations of the set 
$M$.

The spinor inner products~\cite{SpinorHelicity} are denoted by $\spa{i}.j
= \la i^- | j^+\ra$ and $\spb{i}.j = \la i^+| j^-\ra$, where $|i^{\pm}\ra$
are massless Weyl spinors of momentum $k_i$, labeled with the sign of the
helicity.  They are antisymmetric, with norm 
$|\spa{i}.j| = |\spb{i}.j| = \sqrt{s_{ij}}$, where 
$s_{ij} = 2k_i\cdot k_j$, and they carry a relative phase,
$$
{\spb{i}.{j}\over \spa{i}.{j}} 
= - { (k_i^1-ik_i^2)k_j^+ - (k_j^1-ik_j^2)k_i^+ \over
            (k_i^1+ik_i^2)k_j^+ - (k_j^1+ik_j^2)k_i^+ }\ ,
\equn\label{RelativePhase}
$$
where $k_i^+ = k_i^0 + k_i^3$.

Because of the permutation sum, the quantity in square brackets 
in \eqn{BGKMHV} is manifestly symmetric in the labels $2,3,\ldots,n-2$.  
In fact it is symmetric under exchange of all labels $1,2,3,\ldots,n$,
as required by an $N=8$ SWI~\cite{BGK,SusyEight}.  BGK also checked
the behavior of their expression as a graviton momentum becomes soft,
and found the expected universal behavior~\cite{WeinbergSoftG}, 
$$
\M_n^\tree(1,2,\ldots,n^+)\ \mathop{\longrightarrow}^{k_n\to0}\
   {\kappa\over2} \SoftGrav_n \times\ 
   \M_{n-1}^\tree(1,2,\ldots,n-1),
\equn\label{GravTreeSoft}
$$
where the gravitational soft factor for $k_n \rightarrow 0$ is
(for positive helicity) 
$$
\eqalign{
\SoftGrav_n
 &= { -1 \over \spa{1}.{n} \spa{n,}.{n-1} }
  \sum_{i=2}^{n-2} { \spa{1}.{i} \spa{i,}.{n-1} \spb{i}.{n}
                                         \over \spa{i}.{n} }\ .\cr}
\equn\label{GravTreeSoftFactor}
$$
Momentum conservation can be used to show that the soft factor is also 
symmetric under the interchange of legs $1$ and $n-1$ with the others.

In general, one might expect the one-loop generalization of
\eqn{GravTreeSoft} to contain an additional term on the right-hand side,
proportional to a one-loop correction to the soft
factor~(\ref{GravTreeSoftFactor}) multiplied by $\M_{n-1}^\tree$.
However, $\M_{n-1}^\tree$ vanishes for the all-plus helicity
configuration, so such a term cannot appear and the soft behavior of the
one-loop all-plus amplitudes is identical to that of the tree amplitudes,
$$
\M_n^\oneloop(1^+,2^+,\ldots,n^+) 
    \ \mathop{\longrightarrow}^{k_n\to0}\
   {\kappa \over2} \SoftGrav_n \times\ 
   \M_{n-1}^\oneloop(1^+,2^+,\ldots,(n-1)^+).
\equn\label{GravLoopSoft}
$$
More generally, for any helicity configuration one can show that
$\SoftGrav_n$ has no loop corrections to all orders of perturbation
theory; this will be discussed elsewhere~\cite{AllnSusy}.

For the one-loop ansatz, we introduce an off-shell extension of the
BGK tree amplitudes,
$$
\eqalign{
h(a,\{1,2,\ldots,n\},b) &\equiv {\spb1.2 \over \spa1.2}
 { \spab{a}.{\Ksl_{1,2}}.3  \spab{a}.{\Ksl_{1,3}}.4 
   \cdots \spab{a}.{\Ksl_{1,n-1}}.{n}
  \over \spa2.3\spa3.4 \cdots \spa{n-1,}.{n} 
  \, \spa{a}.1 \spa{a}.2\spa{a}.3 \cdots \spa{a}.{n} 
  \, \spa1.{b} \spa{n}.{b} }  \cr
&\hskip1cm + \Perm(2,3,\ldots,n), \cr}
\equn\label{NonRecursiveH}
$$
and $h(a,\{1\},b) \equiv 1/(\spa{a}.{1}^2 \spa{1}.{b}^2)$.
Although it is not obvious in this form, $h$ is symmetric under the 
interchange $a\leftrightarrow b$, and also under the exchange of 
1 with any of the labels $2,3,\ldots,n$.  For example, 
$$
\eqalign{
h(a,\{1,2\},b) &=  
{ \spb1.2 \over 
  \spa1.2 \spa{a}.{1}\spa{1}.{b}\spa{a}.{2}\spa{2}.{b}} \,, \cr
h(a,\{1,2,3\},b) &=
{ \spb1.2\spb2.3\over
  \spa1.2\spa2.3 \spa{a}.{1}\spa{1}.{b}\spa{a}.{3}\spa{3}.{b} }
+ { \spb2.3\spb3.1\over
    \spa2.3\spa3.1 \spa{a}.{2}\spa{2}.{b}\spa{a}.{1}\spa{1}.{b} } \cr
&\hskip1cm
+ { \spb3.1\spb1.2\over
    \spa3.1\spa1.2 \spa{a}.{3}\spa{3}.{b}\spa{a}.{2}\spa{2}.{b}} \,. \cr
}
\equn\label{HalfExamples}
$$
(To obtain the form of $h(a,\{1,2,3\},b)$ in \eqn{HalfExamples} from 
\eqn{NonRecursiveH} requires the Schouten identity,
$\spa{a}.{b}\spa{c}.{d} 
= \spa{a}.{c}\spa{b}.{d} + \spa{a}.{d}\spa{c}.{b}$.)
The symmetry properties of $h$ are manifest for all $n$ in an alternative, 
recursive definition~\cite{AllnSusy}.  

The relation between $h$ and $\M_n^\tree$ is
$$
\eqalign{
{ h(n,\{n-1,n-2,\ldots,2\},1) \over \spa{n}.1^2 } 
  \Bigg|_{k_1+k_2+\cdots+k_n=0}
 &= (-1)^n { \M_n^\tree(1^-,2^-,3^+,\ldots,n^+)
   \over i \, (\kappa/2)^{n-2} \, \spa{1}.{2}^8 }\ . \cr}
\equn\label{hMrelation}
$$
In this form, momentum conservation only has to be used in one factor
in $\M_n^\tree$, in order to convert it into $h$.  Unlike $\M_n^\tree$,
the quantity $h(a,M,b)$ is defined and has simple analytic properties 
even off-shell, i.e. without imposing $k_a+K_M+k_b=0$.  Here $K_M$ is
the sum of the massless momenta in the set $M$,
$K_M^\mu \equiv \sum_{i \in M} k_i^\mu$.

We refer to $h$ as a `half-soft' function, because it satisfies 
$$
h(a,M,b)\ \mathop{\longrightarrow}^{k_m\to0}
\ -\SoftGrav_m(a,M,b) \times h(a,M-m,b),
\hskip2cm \hbox{for $m\in M$}, 
\equn\label{HSoft}
$$
where the half-soft factor
$$
\SoftGrav_m(a,M,b) \equiv { -1 \over \spa{a}.{m}\spa{m}.{b} } 
\sum_{j \in M} 
\spa{a}.{j}\spa{j}.{b} {\spb{j}.{m}\over\spa{j}.{m}} \,
\equn\label{HalfSoftn}
$$
is analogous to the full soft factor~(\ref{GravTreeSoftFactor}),
but its sum is restricted to the set $M$.  \Eqn{HSoft} is easiest to check
from the definition~(\ref{NonRecursiveH}) in the limit $k_1\to0$;
the sum over $(n-1)!$ permutations breaks up into $n-1$ sums of $(n-2)!$
terms, each of which gives $h(a,\{2,3,\ldots,n\},b)$ times a term
in the half-soft factor.

The half-soft functions also solve a recursion relation~\cite{AllnSusy},
$$
h(a,P,b) = - {1\over K_P^2}  
\sum_{ M \cap N = \emptyset \atop M \cup N = P } 
   h(a,M,b) h(a,N,b) \la a^-| \Ksl_M \Ksl_N | a^+\ra
                     \la b^-| \Ksl_M \Ksl_N | b^+\ra \,,
\equn\label{hRecurseqr}
$$
where the sum is over `distinct nontrivial partitions' of $P$ into two
subsets: $M$ and $N$ must both be non-empty, and $(N,M)$ is not distinct
from $(M,N)$.  When $b=a$, this recursion relation arises from considering,
at tree level, off-shell currents for producing a set $P$ of 
identical-helicity gravitons with momenta $k_i$ and polarization tensors 
$\pol_{+}$ which are in a light-cone gauge with respect to $k_a$: 
$\pol_{+}^{\mu\nu} = \spab{a}.{\gamma^\mu}.{i}
\spab{a}.{\gamma^\nu}.{i} /(2\spa{a}.{i}^2)$.  Notice that the effective
vertex $\la a^-| \Ksl_M \Ksl_N | a^+\ra^2$ coincides with that given by
the CS self-dual gravity Lagrangian~(\ref{CSSDG}), 
$(K_M)^{\alpha}{}_{\dot{+}}(K_M)^{\beta}{}_{\dot{+}} 
 (K_N)_{\alpha\dot{+}} (K_N)_{\beta\dot{+}}$, when the fields
$\phi$ carry momentum $K_M$ and $K_N$, and the light-cone direction 
`$\dot{+}$' corresponds to the null-vector $k_a$.  A recursive
construction of the tree-level identical-helicity off-shell gravitational 
current has been given by Rosly and Selivanov~\cite{GravPerturbiner}.

\begin{figure}[ht]
\centerline{\epsfxsize 1.1 truein \epsfbox{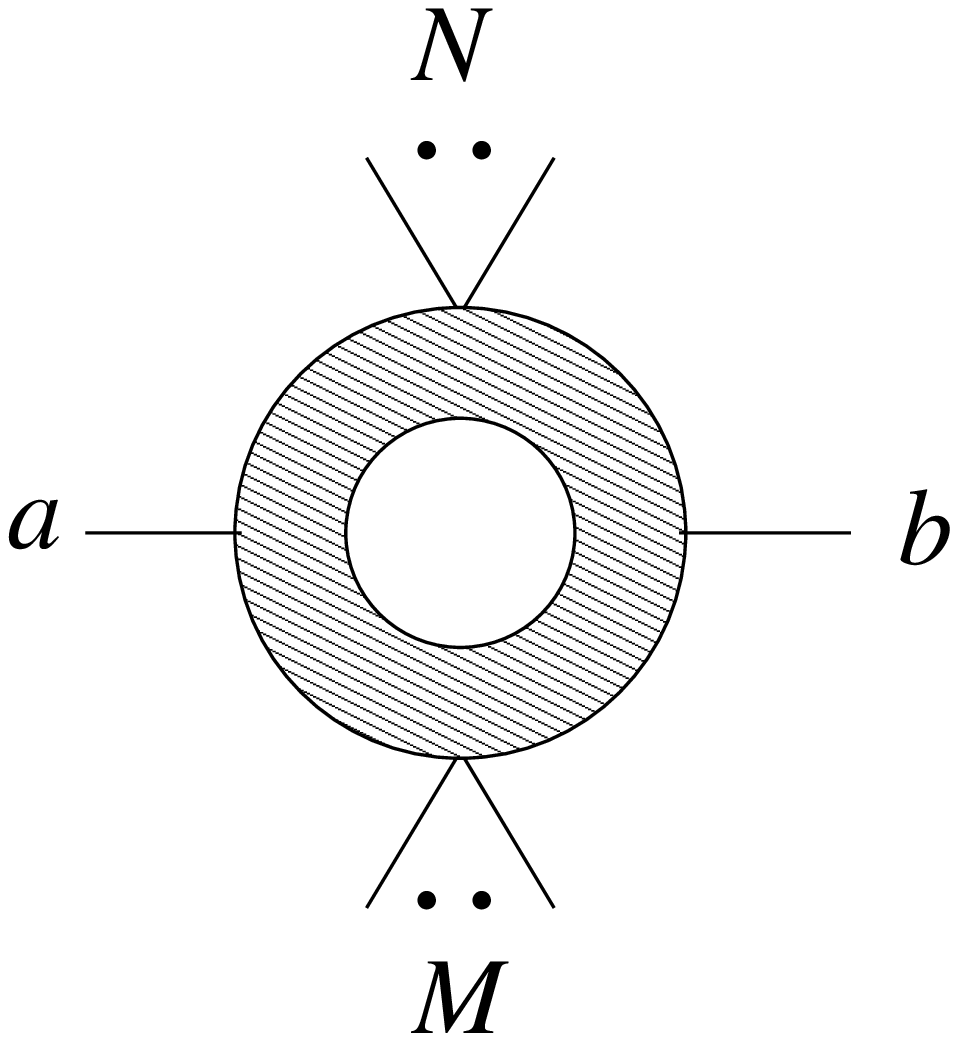}}
\vskip -.2 cm
\caption[]{
\label{LegsFigure}
\small The configurations of external legs that are summed over in 
\eqn{AllPlusSimple}.
}
\end{figure}

By combining products of two half-soft factors which share only their
outside arguments, we can construct a one-loop ansatz which has simple
soft properties.  The ansatz is
$$
\M_n^\oneloop(1^+, 2^+, \ldots, n^+) = 
- {i \over (4 \pi)^2\cdot 960} \, \left({-\kappa\over 2}\right)^n  \,  
\sum_{1 \leq a < b \leq n \atop M, N} h(a, M, b) h(b, N, a) 
 \tr^3[a\, M\, b\, N]\,,
\equn\label{AllPlusSimple}
$$
where $a$ and $b$ are massless legs, and $M$ and $N$ are two sets forming 
a distinct nontrivial partition of the remaining $n-2$ legs, as depicted in 
\fig{LegsFigure}.  Also, 
$\tr[a\, M\, b\, N] \equiv \tr[\ksl_a \Ksl_M \ksl_b \Ksl_N ]$.
For $n=4,5,6$, this ansatz agrees with explicit
computations~\cite{GZ,AllnSusy}, for example
$$
\M_4^\oneloop(1^+,2^+,3^+,4^+) = 
- {i \over (4 \pi)^2 \cdot 120} \, \left({\kappa\over2}\right)^4
 \biggl( {s_{12} s_{23} \over \spa1.2 \spa 2.3 \spa3.4 \spa4.1} \biggr)^2 \,
 (s_{12}^2 + s_{23}^2 + s_{13}^2)\,.
\equn\label{FourPtGrav}
$$

The SWI~(\ref{vanishSWI}), applied to theories with $N=2$ supercharges,
can be used to show~\cite{GZ} that the contribution to the all-plus
amplitude of a massless multiplet of spin $s$ (i.e., two states, with
helicity $\pm s$) circulating in the loop is simply
$$
\M_n^{\oneloop,\ {\rm spin}\ s}(1^+, 2^+, \ldots, n^+)
= (-1)^{2s} \M_n^\oneloop(1^+, 2^+, \ldots, n^+)\,. 
\equn\label{AnyStatesInLoop}
$$

%

The desired soft properties of $\M_n^\oneloop$, \eqn{GravLoopSoft},
follow from those of the half-soft functions, \eqn{HSoft}.  As 
$k_n \to 0$, the $(a,M,b,N)$ term in $\M_{n-1}^\oneloop$ gets
contributions from two terms in $\M_n^\oneloop$, $(a,M+n,b,N)$ and 
$(a,M,b,N+n)$.  Each of the factors $h(a,M+n,b)$ and $h(b,N+n,a)$ 
in \eqn{AllPlusSimple} supplies `half' of the soft factor
in this limit, since
$$
\SoftGrav_n = \SoftGrav_n(a,M,b) + \SoftGrav_n(b,N,a).
\equn\label{ReconSoft}
$$
The trace factors prevent unwanted terms from occurring if $a$ or $b$
becomes soft, but are otherwise innocuous.  


\section{Comparison with All-Plus Gauge Amplitudes}
\label{CompareSection}

The closest analog in gauge theory to the all-plus graviton
amplitude is the one-loop all-plus $n$-gluon amplitude
$\A_n^\oneloop(1^+,2^+,\ldots,n^+)$~\cite{AllPlus,MahlonAllPlus}.
Suppressing the gauge coupling constant $g$ and color factors (they enter
the relation between $\A_n^\oneloop$ and the color-ordered quantity
$A_{n;1}$~\cite{AllPlus}), we have
$$
A_{n;1}(1^+,2^+,\ldots,n^+) = -{i \over 48\pi^2}\,
\sum_{1\leq i_1 < i_2 < i_3 < i_4 \leq n}
{ {\rm tr}_-[i_1 i_2 i_3 i_4]
\over \spa1.2 \spa2.3 \cdots \spa{n}.1 }\ ,
\equn\label{YMAllnPlus}
$$
where $\tr_-[i_1 i_2 i_3 i_4] \equiv {1\over 2}\tr[(1-\gamma_5)
\ksl_{i_1} \ksl_{i_2} \ksl_{i_3} \ksl_{i_4}]$.
Define
$$
g(a,\{1,2,\ldots,n\},b)\ \equiv\ 
{1\over\spa{a}.1\spa1.2\spa2.3\cdots \spa{n-1\ }.{n}\spa{n}.{b}}\ .
\equn\label{gDef}
$$
Just as the identical-helicity graviton current contains $h(a,M,a)$, so
does the identical-helicity gluon current~\cite{RecursiveBG} contain
$g(a,M,a)$.  The $n$-gluon MHV tree amplitude~\cite{ParkeTaylor} can also
be written in terms of $g(a,M,b)$, in an equation similar to
\eqn{hMrelation},
$$
\eqalign{
{ g(1,\{2,3,\ldots,n-1\},n) \over \spa{n}.{1} } 
  \Bigg|_{k_1+k_2+\cdots+k_n=0}
 &= { A_n^\tree(1^-,2^-,3^+,\ldots,n^+) \over i \, \spa{1}.{2}^4 }\ . \cr}
\equn\label{gArelation}
$$
Note that the gauge quantities all have a definite (color) ordering of
their external legs, whereas the corresponding gravitational quantities do
not.

If in the all-plus gauge theory result~(\ref{YMAllnPlus}) we let $i_1 \to a$,
$i_3 \to b$, and rearrange the $i_4$ sum, then we can rewrite the
non-$\gamma_5$ parts of the traces in a form very reminiscent of the
gravity result~(\ref{AllPlusSimple}),
$$
 A_{n;1}(1^+,2^+,\ldots,n^+) \Bigg|_{\hbox{\small non-}\gamma_5} 
= -{i \over (4\pi)^2 \cdot 12} \,
\sum_{1 \leq a < b \leq n \atop M, N} g(a, M, b) g(b, N, a) 
 \tr[a\, M\, b\, N]\,,
\equn\label{AllPlusYMEven}
$$
where $M$ consists of all the legs between $a$ and $b$ (in the cyclic
sense), and $N$ of all the legs between $b$ and $a$.  Unfortunately, a
minus sign prevents us from writing the $\gamma_5$ terms in an analogous
way.  Nevertheless, the similarity between
\eqns{AllPlusSimple}{AllPlusYMEven} is striking, and indeed
\eqn{AllPlusYMEven} helped to motivate the form of the gravitational
ansatz.


\section{Collinear Limits}
\label{ColSection}

As a further consistency check we examine the behavior of the ansatz
(\ref{AllPlusSimple}) as any two external legs become collinear.
Whereas the collinear properties of massless gauge theory amplitudes 
are well-known~\cite{ManganoReview,Review}, the corresponding behavior 
of gravity amplitudes has not been discussed previously in any detail.%
\footnote{The suggestion that collinear limits in gravity are
universal was made by Chalmers and Siegel~\cite{CSUnpublished}.}
The KLT relations provide a simple way to derive the gravity behavior
from that of gauge theory.

In the limit that momenta $k_1$ and $k_2$ are collinear 
($1 \parallel 2$), we have $k_1 \rightarrow z P$, 
$k_2\rightarrow (1-z)P$, where $P = k_1 + k_2$.
Color-ordered tree amplitudes in massless gauge theory satisfy
$$
A_n^\tree(1,2,\ldots, n)
\ \mathop{\longrightarrow}^{1 \parallel 2}\
\sum_{\lambda=\pm}
  \Split^{\rm \qcd\ tree}_{-\lambda}(z,1,2) \times
  A_{n-1}^\tree(P^\lambda, 3,\ldots, n) \,,
\equn\label{YMTreeColl}
$$
where the sum runs over the two helicities of the fused leg.  The
splitting amplitudes $\Split^{\rm \qcd\ tree}_{-\lambda}(z,1,2)$ are
universal: they depend only on the momenta and helicities of the legs
becoming collinear, and not upon the specific amplitude under
consideration.  Here we are interested in the pure gluon splitting
ampitudes with both legs of positive helicity,
$$
\eqalign{
& \Split_+^{\rm \qcd\ tree}(z, 1^+, 2^+) = 0\,,\cr
& \Split_-^{\rm \qcd\ tree}(z, 1^+, 2^+) = {1\over \sqrt{z (1-z)}} 
       {1\over \spa1.2} \,.\cr}
\equn\label{YMTreeSplit}
$$

The KLT relations for $n=4,5$, in the limit of infinite string tension, 
read~\cite{KLT,BGK}
$$
\eqalign{
\M_4^\tree (1,2,3,4) & = 
- i \left({\kappa\over2}\right)^2 s_{12} A_4^\tree(1,2,3,4) \,
                 A_4^\tree(1,2,4,3)\,, \cr
\M_5^\tree(1,2,3,4,5) & = 
i \left({\kappa\over2}\right)^3 s_{12} s_{34}  A_5^\tree(1,2,3,4,5) \, 
                   A_5^\tree(2,1,4,3,5)\ +\ \Perm(2,3) \,. \cr}
\equn\label{KLTRel}
$$
Because the gauge theory amplitudes on the right-hand side of \eqn{KLTRel}
(and similar equations for $n>5$) have the universal collinear 
behavior~(\ref{YMTreeColl}), the gravity amplitudes must obey,
$$
\M_n^\tree(1,2,\ldots, n)
\ \mathop{\longrightarrow}^{1 \parallel 2}\
 {\kappa\over2} \sum_{\lambda=\pm}
  \SplitGrav_{-\lambda}(z,1,2) \times
   \M_{n-1}^\tree(P^\lambda, 3,\ldots, n) \,,
\equn\label{GravTreeColl}
$$
with splitting amplitudes
$$
\eqalign{
\SplitGrav_{-(\lambda+\tilde\lambda)}
(z,1^{\lambda_1+\tilde\lambda_1},2^{\lambda_2+\tilde\lambda_2})\ =\  
-s_{12} &\times \Split_{-\lambda}^{\rm \qcd\ tree}
          (z,1^{\lambda_1},2^{\lambda_2}) \cr
        &\times \Split_{-\tilde\lambda}^{\rm \qcd\ tree}
          (z,2^{\tilde\lambda_2},1^{\tilde\lambda_1})\,. \cr}
\equn\label{GravYMSplit}
$$
Inserting, for example, the values of the pure gluon splitting 
amplitudes~(\ref{YMTreeSplit}) into \eqn{GravYMSplit}, gives
$$
\eqalign{
 \SplitGrav_{+}(z,1^{+},2^{+}) &= 0\,,\cr
 \SplitGrav_{-}(z,1^{+},2^{+})
            &= { -1 \over z(1-z) }
                 { \spb{1}.{2} \over \spa{1}.2 }\ .\cr}
\equn\label{GravTreeSplit}
$$

%
\begin{figure}[ht]
\centerline{\epsfxsize 1.5 truein \epsfbox{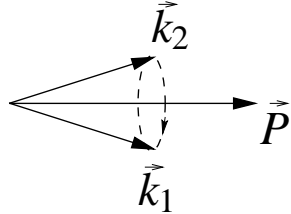}}
\vskip -.2 cm
\caption[]{
\label{RotateColFigure}
\small As two momenta become collinear the gravity $S$-matrix
develops a phase singularity which can be detected by rotating the 
two momenta about the axis formed by their sum.
}
\end{figure}

The meaning of \eqn{GravTreeColl} is slightly different from
the gauge theory case.  In the gauge case, the leading power-law behavior 
of the amplitude in the collinear limit is determined by the universal 
terms~(\ref{YMTreeColl}); non-universal terms are suppressed by a 
relative power of $\sqrt{s_{12}}$.  In the case of 
\eqns{GravTreeColl}{GravTreeSplit},
there are non-universal terms of the same magnitude as 
$\spb{1}.{2}/\spa{1}.2$ as $s_{12} \to 0$.  However, these terms 
do not acquire any {\it phase} as the nearly collinear three-vectors 
$\vec k_1$ and $\vec k_2$ are rotated around their sum $\vec P$.
For example, consider the two factors,
$$
\hbox{(a)}: \quad {\spb1.2 \over \spa1.2} \,, \hskip 2 cm 
\hbox{(b)}: \quad {\spb1.3 \over \spa1.3} \,. 
$$
If we take $\vec k_1$ to be nearly collinear with $\vec k_2$ and rotate
$\vec k_1$ and $\vec k_2$ around the vector $\vec P= \vec k_1 + \vec k_2$
by a large angle, as depicted in \fig{RotateColFigure}, the factor (b)
undergoes only a slight numerical variation.  On the other hand, from
\eqn{RelativePhase}, the factor (a) will undergo a large phase variation
depending on the angle of rotation.  (As $\vec k_1$ and $\vec k_2$ rotate
once around their sum $\vec P$, the phase of
$\SplitGrav_{-}(z,1^{+},2^{+})$, or of the factor (a), changes by $4\pi$, due
to the angular-momentum mismatch of 2$\hbar$ between the graviton $P^+$
and the pair of gravitons $(1^+,2^+)$.)  One may therefore numerically
separate the terms with phase singularities, by evaluating the amplitude
for $\vec k_1$ and $\vec k_2$ almost collinear, and subtracting the same
expression but with the two collinear momenta rotated by a large angle
about $\vec P$; the non-universal terms will then cancel.
(In signature $(2,2)$, the spinor products $\spa1.2$ and $\spb1.2$ are
not complex conjugates of each other; thus $\spa1.2$ can be taken to zero 
independently of $\spb1.2$, in order to separate out the universal
terms~\cite{CSUnpublished}.)

For the one-loop all-plus amplitudes, any loop corrections to the 
splitting amplitudes must drop out, because they would be multiplied 
by vanishing tree amplitudes, according to \eqn{vanishSWI}.  
Thus we require in the collinear limit,
$$
\M_{n}^{\oneloop}(1^+, 2^+, 3^+, \ldots, n^+)
\ \mathop{\longrightarrow}^{1 \parallel 2}\
 {\kappa\over2} \sum_{\lambda=\pm}  
  \SplitGrav_{-\lambda} (z,1^+,2^+)\,
  \times \M_{n-1}^{\oneloop}(P^\lambda, 3^+,\ldots,n^+)  \,.
\equn\label{GravLoopSplit}
$$
(Actually, for any helicity configuration the gravity splitting amplitudes,
like the soft factors, do not have corrections at any loop 
order~\cite{AllnSusy}.)

To verify the behavior~(\ref{GravLoopSplit}), two different types of 
limits of the half-soft functions are needed,
$$
\eqalign{
h(a,\{1,2,3,\ldots,n\},b)\ &\mathop{\longrightarrow}^{1 \parallel 2}\ 
 {1\over z(1-z)} {\spb1.2\over\spa1.2} 
 \times h(a,\{P,3,\ldots,n\},b) \,, \cr
h(1,\{2,3,\ldots,n\},b)\ &\mathop{\longrightarrow}^{1 \parallel 2}\ 
  {1 \over \spa{1}.{2}} \, 
{ \spa{1}.{b} \langle b^-| \Ksl_{3,n} | 2^-\rangle 
  \over \spa{2}.{b}^2 } \times h(1,\{3,\ldots,n\},b) \,, \cr}
\equn\label{HCol}
$$
where we have used the Schouten identity, and dropped terms without phase
singularities.  Note that the second, more singular limit reduces to the
same degree of singularity in the on-shell case, $K_{1,n}+k_b = 0$ --- as
it must in order for eqs.~(\ref{HCol}) to give the correct collinear
limits in all channels of the MHV tree amplitudes~(\ref{BGKMHV}), via the
relation~(\ref{hMrelation}).  Using eqs.~(\ref{HCol}) it is
straightforward to establish that \eqn{AllPlusSimple} has the correct
phase singularities as any two momenta become collinear.


\section{Comments}
\label{CommentSection}

Although the analytic requirements that we have imposed on the amplitudes
should be sufficiently stringent to uniquely fix them, it would still be
useful to have a complete proof that the expression~(\ref{AllPlusSimple})
is correct.  The ansatz for the all-plus gauge amplitudes~\cite{AllPlus}
was proven using a recursive construction of a doubly-off-shell current in
gauge theory~\cite{MahlonAllPlus}.  We have constructed the analogous
doubly-off-shell currents in gravity, and have sewn them into the all-plus
gravity amplitudes for $n=4,5$, obtaining numerical agreement with
\eqn{AllPlusSimple}~\cite{AllnSusy}.  However, we have not been able to
analytically simplify the resulting expressions for general $n$.

In this paper we have found an infinite sequence of {\it non}-vanishing 
one-loop amplitudes for identical-helicity gravitons in Minkowski space
(signature $(1,3)$).  (Although the $n>6$ results are, strictly speaking, 
an ansatz, the structure of the soft and collinear limits certainly 
requires these amplitudes to be nonzero, given that the lower-point
amplitudes are.)  In contrast, all one-loop scattering
amplitudes have been argued to vanish for the closed $N=2$
string~\cite{BerkVafa}, which is supposed to describe self-dual gravity in
signature $(2,2)$.%
\footnote{However, two explicit computations claim to have found a nonvanishing
one-loop three-point amplitude~\cite{BGIK}.}  The same paradoxical 
situation has been noted for SDYM amplitudes and the open $N=2$ 
string~\cite{ChalmersSiegel}.  

It has been argued that the effective actions for both closed and open
$N=2$ strings are modified upon including the effects of instantons for
the $U(1)$ world-sheet current~\cite{LechtSiegel}.  In the case of the
closed $N=2$ string, originally thought~\cite{NTwoStringOV} to be described
by the `Pleba\'nski' Lagrangian~\cite{SDG},
$$
{\cal L}_{\rm P}^{\rm SDG}
 =  \phi \Bigl( {1\over2} \del^2 \phi 
   + {\kappa\over3}(\del^{\alpha}{}_{\dot{+}}\del^{\beta}{}_{\dot{-}}\phi)
                   (\del_{\alpha\dot{+}}\del_{\beta\dot{-}}\phi) \Bigr) \,,
\equn\label{PlebSDG}
$$
this Lagrangian is to be converted into the CS SDG Lagrangian~(\ref{CSSDG})
(or a one-field version of it)~\cite{LechtSiegel}.  The difference between
\eqns{PlebSDG}{CSSDG} might account for the different conclusions that
have been reached about the vanishing of one-loop self-dual amplitudes.
Clearly, a direct comparison of amplitudes obtained from the two actions in
signature $(2,2)$ would be required to demonstrate this.

Another, possibly related, explanation involves a potential anomaly in the
string world-sheet theory~\cite{ChalmersSiegel}.  We have nothing to say
about the string theory situation.  However, Bardeen has suggested that 
the non-vanishing of the $(1,3)$ SDYM one-loop amplitudes is related to 
an anomaly in the SDYM conserved currents~\cite{Bardeen}.  In this
context, we note that the existing $(1,3)$ Minkowski-space field theory 
calculations used regularizations which do not respect self-duality, 
even though the final result is finite.
The unitarity-cut calculation requires $D \neq 4$, for example, in order
to detect the rational functions constituting the result.  Mahlon's
recursive Feynman-diagram approach to the all-plus gauge amplitudes also
used dimensional regularization; without it, naive manipulations would
have given zero as the answer~\cite{MahlonAllPlus}.

In addition to self-dual gravity and Yang-Mills theory, it is possible to
couple the two theories to each other, as in the $N=2$ open and heterotic
strings~\cite{NTwoOpenHetString}.  Off-shell currents and MHV amplitudes
for a mixture of gravitons and gauge bosons of identical helicity have 
been constructed at tree level~\cite{SelivanovMixed}.  The one-loop all-plus 
gravity/gauge amplitudes also have simple general analytic properties, 
and it would not be surprising if they too could be determined in closed form.


\vskip .3 cm 
\noindent
{\bf Acknowledgments}

We thank G. Chalmers, D.C. Dunbar, A.K. Grant and D.A. Kosower for
helpful discussions.


\end{document}